\documentclass[aps,prd,reprint,nofootinbib,longbibliography,superscriptaddress]{revtex4-2}

\usepackage[T1]{fontenc}
\usepackage[utf8]{inputenc}
\usepackage{lmodern}
\usepackage{amsmath,amssymb,amsthm,mathtools,bm}
\usepackage{physics}
\usepackage{microtype}
\usepackage{hyperref}
\usepackage{xcolor}

\hypersetup{colorlinks=true,citecolor=blue,linkcolor=blue,urlcolor=blue}

\newcommand{\Afun}{\mathcal{A}}
\newcommand{\Bfun}{\mathcal{B}} 
\newcommand{\Vfun}{\mathcal{V}}
\newcommand{\Ione}{\mathcal{I}_1}
\newcommand{\Itwo}{\mathcal{I}_2}
\newcommand{\Ithree}{\mathcal{I}_3}
\newcommand{\Ifive}{\mathcal{I}_5}
\newcommand{\eh}{\hat{g}}
\newcommand{\gJ}{\tilde{g}}
\newcommand{\nJ}{\tilde{\nabla}}
\newcommand{\RJ}{\tilde{R}}
\newcommand{\BoxJ}{\tilde{\Box}}
\newcommand{\The}{\Theta}
\newcommand{\DD}{\mathrm{D}}
\newcommand{\chiJ}{\chi_{\mathrm J}}
\newcommand{\VJ}{\mathcal{U}}
\newcommand{\omegaI}{\omega_{\mathrm{inv}}}

\begin{document}

\title{Jordan-like invariant representations of scalar-tensor gravity and the return of imperfect-fluid thermodynamics}

\author{David S. Pereira}
\email{djpereira@ciencias.ulisboa.pt}

\author{Jos\'e Pedro Mimoso}
\email{jpmimoso@ciencias.ulisboa.pt}

	\affiliation{%
		Departamento de F\'{i}sica, Faculdade de Ci\^{e}ncias da Universidade de Lisboa, Campo Grande, Edif\'{\i}cio C8, P-1749-016 Lisbon, Portugal}
	\affiliation{Instituto de Astrof\'{\i}sica e Ci\^{e}ncias do Espa\c{c}o, Faculdade de
		Ci\^encias da Universidade de Lisboa, Campo Grande, Edif\'{\i}cio C8,
		P-1749-016 Lisbon, Portugal;\\
	}%

\date{\today}

\begin{abstract} 
It has been shown that the thermodynamics of frame-invariant scalar-tensor gravity admits an Einstein-frame-like invariant representation, in which the scalar sector is minimally coupled and behaves as a perfect fluid with vanishing temperature. We show that this is not the only invariant thermodynamic organization available. Using instead the invariant matter metric, we construct a Jordan-frame-like invariant representation in which matter is minimally coupled and the gravitational action takes the Brans-Dicke-Bergmann-Wagoner form with invariant fields $\Psi=\mathcal I_1^{-1}$, $\mathcal U(\Psi)$, and $\omega_{\mathrm{inv}}(\Psi)$. In this representation the metric field equations retain the Hessian sector characteristic of Jordan-frame scalar-tensor gravity. A $1+3$ decomposition of the corresponding effective stress tensor yields nonvanishing heat flux and anisotropic stress in a generic congruence. In the $\Psi$-comoving frame, the usual first-order thermodynamic identifications are recovered in invariant form, including $K_{\mathrm J}T_{\mathrm J}=-\dot\Psi/(\kappa^2\Psi)$ and $\eta_{\mathrm J}=\dot\Psi/(2\kappa^2\Psi)$, with the bulk channel obtained in homogeneous and isotropic sectors. Thus, frame-invariant scalar-tensor thermodynamics is not intrinsically restricted to an Einstein-frame-like perfect-fluid description: it also admits a Jordan-frame-like invariant formulation in which the effective imperfect-fluid interpretation is manifest. \end{abstract}

\maketitle

\section{Introduction}

Scalar-tensor theories constitute one of the simplest and most extensively studied extensions of general relativity. Their defining feature is the presence of, in addition to the metric tensor, one or more scalar degrees of freedom that participate in the gravitational interaction. The original motivations go back to the Jordan--Brans--Dicke idea of promoting the gravitational coupling to a dynamical field~\cite{Jordan:1959eg,Brans:1961sx}, and were subsequently generalized to the broader Bergmann--Wagoner class of scalar-tensor theories~\cite{Bergmann:1968ve,Wagoner:1970vr}. Scalar degrees of freedom also arise naturally in dimensional reductions, low-energy limits of high-energy theories, inflationary model building, dark-energy phenomenology, early Universe phenomena and effective descriptions of modified gravity~\cite{Fujii:2003pa,Faraoni:2004pi,Clifton:2011jh,Pereira:2024ddu,Pereira:2025rsr,CosmoVerseNetwork:2025alb,CANTATA:2021asi,Sotiriou:2010zz}. For this reason scalar-tensor gravity provides both a theoretically motivated and phenomenologically flexible framework for parametrizing deviations from general relativity.

A characteristic feature of scalar-tensor gravity is that the same theory can be written in different conformal parametrizations. The most familiar examples are the Jordan and Einstein frames. In a Jordan parametrization, matter is minimally coupled to the metric, while the scalar field typically couples nonminimally to the Ricci scalar. In an Einstein parametrization, the gravitational part of the action is brought to the Einstein-Hilbert form, but matter is generally coupled to a scalar-dependent conformal metric~\cite{Fujii:2003pa,Faraoni:2004pi}. These parametrizations are related by conformal transformations of the metric together with redefinitions of the scalar field. Consequently, a central question is not merely how a given equation looks in a chosen frame, but which statements have a meaning independent of that choice~\cite{Damour:1992we,Jarv:2014hma}.

The invariant approach addresses precisely this issue. Instead of treating a particular conformal frame as fundamental, one constructs combinations of the metric, scalar field, and model functions that remain unchanged under simultaneous Weyl rescalings and scalar-field reparametrizations~\cite{Jarv:2014hma}. In this language, a ``frame'' is a parametrization of the same underlying scalar-tensor theory, whereas an invariant quantity is one whose value and functional meaning do not depend on which parametrization is used. This distinction is especially important for effective thermodynamic interpretations of scalar-tensor gravity~\cite{Faraoni:2021jri,Faraoni:2022ixc,Faraoni:2025alq,Banerjee:2026ulx,Faraoni:2025ufi,Faraoni:2023hwu,Giardino:2022sdv,Faraoni:2022doe,Faraoni:2022gry,Giardino:2023ygc,Giusti:2021sku,Faraoni:2021lfc,Faraoni:2025fjq,Faraoni:2025dex,Gallerani:2025myd,Pereira:2026xog,Jarv:2026dbb,Banerjee:2026ulx,Faraoni:2026wlc,Pereira:2025dmk,Gallerani:2024gdy,Pereira:2026cum}, because quantities such as heat flux, anisotropic stress, temperature, or chemical-potential-like variables can depend on how the scalar sector is organized.

A particularly important choice is the invariant metric. The Einstein-like invariant representation $\eh_{ab}=\Afun g_{ab}$ brings the gravitational action into a minimally coupled form and is therefore especially convenient for comparing scalar-tensor gravity with Einstein gravity. However, as emphasized already in Ref.~\cite{Jarv:2014hma}, this invariant metric is not unique. Other invariant metrics may be chosen, and different choices can make different structural features of the same theory manifest.

The purpose of this paper is to construct and analyze a Jordan-like invariant representation. The representation is defined by the invariant matter metric
\begin{equation}
\gJ_{ab}=e^{2\alpha(\Phi)}g_{ab},
\end{equation}
where $e^{2\alpha(\Phi)}g_{ab}$ is the metric appearing in the matter action $S_m[e^{2\alpha(\Phi)}g_{ab},\chi]$. Matter is therefore minimally coupled to $\gJ_{ab}$ by construction. When the theory is written in terms of this metric and the invariant scalar $\Psi$, the action takes the Brans-Dicke-Bergmann-Wagoner form with invariant functions $\mathcal U(\Psi)$ and $\omega_{\mathrm{inv}}(\Psi)$ and the resulting field equations contain the Hessian sector familiar from Jordan-frame scalar-tensor gravity, and this sector generates an effective imperfect-fluid decomposition.

This construction is motivated in part by the recent invariant thermodynamic analysis of Ref.~\cite{Jarv:2026dbb}, which focuses on the Einstein-like invariant representation. In that representation the scalar sector is minimally coupled and has the stress tensor of a perfect fluid; the associated invariant temperature therefore vanishes, and the departure from general relativity is described instead by a chemical-potential or diffusive variable, in analogy with minimally coupled scalar fields~\cite{Faraoni:2022ixc}. We do not dispute this conclusion within the Einstein-like representation. Rather, we clarify its scope by showing that the invariant formalism also admits a Jordan-like representation in which the same scalar-tensor dynamics are organized as an effective imperfect fluid leading to a thermodynamic interpretation instead of a hydrodynamical.

The main result is therefore a statement about representation dependence. The Einstein-like and Jordan-like invariant metrics are both mathematically legitimate representations of the same conformal class, but they organize the scalar sector differently. In the former, the scalar sector is naturally described as a minimally coupled perfect fluid with vanishing temperature; in the latter, the nonminimal-coupling Hessian terms are explicit and the usual temperature-based imperfect-fluid interpretation is recovered.

This paper is organized as follows. In Sec.~\ref{sec:prelim} we review the invariant variables of scalar-tensor gravity and recall the Einstein-like invariant representation used in recent discussions of frame-invariant thermodynamics. In Sec.~\ref{sec:jmetric} we introduce a Jordan-like invariant metric, built from the invariant matter metric, and define the corresponding invariant coupling field and potential. Section~\ref{sec:action} derives the action in this Jordan-like invariant representation, introduces the invariant Brans-Dicke function, and shows that the action takes the Brans-Dicke-Bergmann-Wagoner form while remaining fully invariant. In Sec.~\ref{sec:field_eqs} we obtain the associated field equations and identify the Hessian sector responsible for the effective imperfect-fluid structure. The $1+3$ decomposition is carried out in Sec.~\ref{sec:1plus3}, where the invariant heat flux and anisotropic stress are displayed explicitly. In Sec.~\ref{sec:thermo} we specialize to the scalar-comoving frame and recover the usual first-order thermodynamic identifications in invariant Jordan-like variables. Section~\ref{sec:comparison} compares the Einstein-like and Jordan-like invariant representations, emphasizing that they provide different but equally invariant organizations of the same scalar-tensor dynamics. We conclude in Sec.~\ref{sec:discussion} with a summary and possible extensions.

\section{Invariant preliminaries}
\label{sec:prelim}

We begin from the standard scalar-tensor action with one scalar field~\cite{Jarv:2014hma}
\begin{align}
S&=\frac{1}{2\kappa^2}\int d^4x\sqrt{-g}\,\Big[\Afun(\Phi)R-\Bfun(\Phi)g^{ab}\nabla_a\Phi\nabla_b\Phi \nonumber\\
&-2\ell^{-2}\Vfun(\Phi)\Big]+S_m\!\left[e^{2\alpha(\Phi)}g_{ab},\chi\right],
\label{eq:general_action}
\end{align}
where $\kappa^2=8\pi G$, $\ell$ is a constant length scale, and $\chi$ collectively denotes matter degrees of freedom. The four functions $\Afun(\Phi)$, $\Bfun(\Phi)$, $\Vfun(\Phi)$, and $\alpha(\Phi)$ specify a particular parametrization of the theory~\cite{Jarv:2014hma}. The function $\Afun$ controls the nonminimal coupling between the scalar field and curvature, $\Bfun$ determines the normalization of the scalar kinetic term, $\Vfun$ is the scalar potential, and $\alpha$ determines the metric to which matter is minimally coupled. In the usual Jordan parametrization one sets $\alpha=0$, so that matter is minimally coupled to $g_{ab}$, whereas in an Einstein parametrization one chooses variables in which the gravitational Ricci term has the Einstein-Hilbert normalization~\cite{Fujii:2003pa,Faraoni:2004pi}. These are not, in general, different theories, but different coordinate choices on the space of scalar-tensor representations.

For definiteness, we use the same convention as in Ref.~\cite{Jarv:2014hma}: a change of scalar-tensor parametrization is written as
\begin{equation}
g_{ab}=e^{2\gamma(\bar\Phi)}\bar g_{ab},
\qquad
\Phi=f(\bar\Phi),
\label{eq:frame_transformation}
\end{equation}
where $\gamma$ and $f$ are arbitrary sufficiently regular functions, with $f'(\bar\Phi)\neq0$ on the branch considered. The action retains the form \eqref{eq:general_action} provided the model functions transform as
\begin{align}
\bar{\Afun}(\bar\Phi)
&=e^{2\gamma(\bar\Phi)}\Afun(f(\bar\Phi)),
\\
\bar{\Vfun}(\bar\Phi)
&=e^{4\gamma(\bar\Phi)}\Vfun(f(\bar\Phi)),
\\
\bar{\alpha}(\bar\Phi)
&=\alpha(f(\bar\Phi))+\gamma(\bar\Phi),
\\
\bar{\Bfun}(\bar\Phi)
&=e^{2\gamma(\bar\Phi)}
\bigg[
\big(f'(\bar\Phi)\big)^2\Bfun(f(\bar\Phi))
\nonumber\\
&\qquad
-6f'(\bar\Phi)\gamma'(\bar\Phi)\Afun'(f(\bar\Phi))
-6\big(\gamma'(\bar\Phi)\big)^2\Afun(f(\bar\Phi))
\bigg].
\label{eq:function_transformations}
\end{align}
Here a prime on $f$ or $\gamma$ denotes differentiation with respect to $\bar\Phi$, whereas a prime on $\Afun$ denotes differentiation with respect to its own argument. Invariance in this paper refers to invariance under the combined transformation \eqref{eq:frame_transformation}--\eqref{eq:function_transformations}, not to spacetime diffeomorphism invariance, which is already built into the covariant action. With these transformation rules, one verifies directly that
\begin{equation}
\bar{\Ione}=\Ione,\qquad
\bar{\Itwo}=\Itwo,\qquad
\bar{\Ithree}=\Ithree,
\end{equation}
up to the usual branch sign and additive constant in $\Ithree$. Moreover, one has that
\begin{equation}
\bar{\eh}_{ab}
=
\bar{\Afun}\bar g_{ab}
=
\Afun g_{ab}
=
\eh_{ab},
\end{equation}
and
\begin{equation}
\bar{\gJ}_{ab}
=
e^{2\bar\alpha}\bar g_{ab}
=
e^{2\alpha}g_{ab}
=
\gJ_{ab},
\end{equation}

Thus both $\eh_{ab}$ and $\gJ_{ab}$ are invariant metric representations of the same scalar-tensor conformal class. This point is conceptually important for the present work. If two expressions differ only because different frame variables have been chosen, then they should not be interpreted as describing distinct physics. Conversely, if a statement depends on which invariant metric representation is used, then it is not a statement about the invariant formalism alone, but about a particular invariant organization of the conformal class. The distinction between frame dependence and representation dependence is therefore essential in comparing the Einstein-like and Jordan-like invariant descriptions.

Under a conformal transformation of the metric and a scalar-field reparametrization, the model functions change, but certain combinations remain invariant. A convenient basis of scalar invariants is~\cite{Jarv:2014hma}
\begin{equation}
\Ione\equiv \frac{e^{2\alpha}}{\Afun},
\quad
\Itwo\equiv \frac{\Vfun}{\Afun^2},
\quad
\Ithree\equiv \pm\int d\Phi\,\sqrt{\frac{2\Afun\Bfun+3(\Afun')^2}{4\Afun^2}}.
\label{eq:I123}
\end{equation}
Here and below a prime denotes derivative with respect to $\Phi$. The quantity $\Ithree$ is defined up to an additive constant and branch sign, as usual for an antiderivative. The invariant $\Ione$ measures the relative conformal coupling between the metric used in the gravitational part of the action and the metric seen by matter~\cite{Jarv:2014hma}. Equivalently, it encodes the conformal factor relating the Einstein-like invariant metric to the matter metric. The invariant $\Itwo$ is the scalar potential expressed in units of the squared nonminimal coupling, and therefore gives the potential naturally associated with the Einstein-like invariant representation. The invariant $\Ithree$ defines a frame-independent scalar-field distance: it is the scalar variable that canonically normalizes the kinetic sector when the theory is written in the Einstein-like invariant metric~\cite{Jarv:2014hma}.

These invariants are useful as they separate two issues that are often conflated. The first is the choice of parametrization, such as Jordan or Einstein variables. The second is the choice of invariant representation, namely which invariant metric is used to organize the same conformal class. The combinations $\Ione$, $\Itwo$, and $\Ithree$ remove the first ambiguity, since they can be computed in any parametrization and give the same invariant content~\cite{Jarv:2014hma}. However, they do not by themselves select a unique invariant metric. One may use the Einstein-like representation $\eh_{ab}=\Afun g_{ab}$, which makes the gravitational sector minimal, or one may use the matter metric $\gJ_{ab}=e^{2\alpha}g_{ab}$, which keeps the matter coupling minimal. Both are invariant, but they emphasize different physical structures.

This observation is crucial for the thermodynamic interpretation developed in this work. J\"arv \textit{et al.}\ further define the Einstein-like invariant metric~\cite{Jarv:2014hma}
\begin{equation}
\eh_{ab}\equiv \Afun\,g_{ab},
\label{eq:ghat_def}
\end{equation}
which is invariant under the combined frame transformation. In terms of $\eh_{ab}$ and the invariants \eqref{eq:I123}, the action can be written, up to the standard boundary term discussed in Ref.~\cite{Jarv:2014hma}, as
\begin{align}
S&=\frac{1}{2\kappa^2}\int d^4x\sqrt{-\eh}\,\Big[\hat R-2\eh^{ab}\hat\nabla_a\Ithree\hat\nabla_b\Ithree\nonumber\\
&-2\ell^{-2}\Itwo\Big]+S_m\!\left[\Ione\eh_{ab},\chi\right].
\label{eq:einvariant_action}
\end{align}
The recent thermodynamic analysis of Ref.~\cite{Jarv:2026dbb} focuses precisely on the minimally coupled gravitational part of \eqref{eq:einvariant_action}. Since the scalar sector is minimal in this representation, its stress tensor is that of a perfect fluid, and the corresponding invariant temperature vanishes identically, exactly as in the minimal-scalar analysis of Ref.~\cite{Faraoni:2022ixc}. This motivates asking how the effective-fluid interpretation changes when one chooses a different invariant metric representation.

\subsection{A Jordan-like invariant metric}
\label{sec:jmetric}

The crucial observation, already emphasized in Ref.~\cite{Jarv:2014hma}, is that the invariant metric choice is not unique. This point is central for the interpretation of recent invariant thermodynamic claims. If the invariant formalism admitted only the Einstein-like metric, then the perfect-fluid, zero-temperature description would indeed be forced by the formalism itself. But once the nonuniqueness of the invariant metric is taken seriously, this conclusion no longer follows. Different invariant representations reorganize the same underlying scalar-tensor theory in different ways, and it becomes a substantive question whether the perfect-fluid description is truly unique or merely one particularly convenient invariant packaging. 
Since $\Ione$ is itself invariant, the product
\begin{equation}
\gJ_{ab}\equiv \Ione\,\eh_{ab}=e^{2\alpha}g_{ab}
\label{eq:gJ_def}
\end{equation}
is equally invariant. This metric has a distinguished property: the matter action is minimally coupled to it by construction,
\begin{equation}
S_m\!\left[e^{2\alpha}g_{ab},\chi\right]=S_m\!\left[\gJ_{ab},\chi\right],
\label{eq:Sm_minimal_tilde}
\end{equation}
hence, for this reason, we shall call $\gJ_{ab}$ the \emph{Jordan-like invariant metric}. It is not unique either, but it is the natural invariant representation that coincides with the physical matter metric.

A qualification concerning the role of matter in selecting this invariant representative is important. There is a distinction between matter being dynamically negligible and the matter sector being absent from the theory altogether. If the theory contains the matter action $S_m[e^{2\alpha(\Phi)}g_{ab},\chi]$ but a particular solution or cosmological epoch has $T^{(m)}_{ab}=0$---for instance because ordinary matter has been diluted to negligible density during an inflationary stage---the coupling function $\alpha(\Phi)$ and hence the invariant metric $\gJ_{ab}=e^{2\alpha}g_{ab}$ remain well defined as part of the underlying theory. The Jordan-like invariant representation can therefore be used without modification even though matter does not contribute dynamically to that particular solution.

If, instead, the matter sector is removed from the theory altogether, so that $S_m$ is identically absent, no physical matter metric is available to single out $\gJ_{ab}$. In that genuinely vacuum theory one may still choose an invariant conformal representative having the Jordan-like BDBW form, for example by retaining $\alpha$ as a fiducial continuation from an underlying matter-coupled theory, but this representative is then no longer distinguished by matter coupling. Correspondingly, the thermodynamic quantities associated with such a choice should be understood as belonging to that invariant representation rather than as being selected by a physical matter sector.

The terminology ``Jordan-like'' is justified by the following checks.

\paragraph*{Jordan frame.}
In a Jordan parametrization with $\alpha=0$, Eq.~\eqref{eq:gJ_def} reduces to
\begin{equation}
\gJ_{ab}=g_{ab}.
\end{equation}
Therefore the invariant Jordan-like metric coincides with the usual Jordan metric.

\paragraph*{Einstein frame.}
In a canonical Einstein parametrization with $\Afun=1$, one has $\eh_{ab}=g^{\rm E}_{ab}$, while
\begin{equation}
\gJ_{ab}=e^{2\alpha_{\rm E}}g^{\rm E}_{ab}.
\end{equation}
Thus the same invariant object is represented in Einstein variables by a conformal rescaling of the Einstein metric.

\paragraph*{Invariant coupling field.}
Because $\Ione$ is invariant, so is its inverse. We therefore define the invariant Jordan scalar
\begin{equation}
\Psi\equiv \Ione^{-1}=\Afun\,e^{-2\alpha},
\label{eq:Psi_def}
\end{equation}
that reduces to the usual Jordan coupling function in a Jordan parametrization. It is the natural invariant analogue of the Brans-Dicke field.

\paragraph*{Invariant potential}
On any branch for which $\Psi(\Phi)$ is locally invertible, we may regard invariant functions of $\Phi$ as functions of $\Psi$. In particular,
\begin{equation}
\VJ(\Psi)\equiv \frac{\Itwo}{\Ione^2}=\Vfun\,e^{-4\alpha},
\label{eq:Uinv_def}
\end{equation}
is the potential measured in the Jordan-like invariant representation. 

\section{Rewriting the action in Jordan-like invariant form}
\label{sec:action}

It is useful to derive the Jordan-like invariant action directly from the original scalar-tensor action~\eqref{eq:general_action}, rather than from the Einstein-like invariant action~\eqref{eq:einvariant_action}. This makes clear that the Jordan-like representation is not obtained merely by reversing the Einstein-like construction, but arises directly once one chooses the invariant matter metric as the basic geometrical variable. An equivalent derivation can of course be carried out starting from Eq.~\eqref{eq:einvariant_action}; we comment on this at the end of the section.

Since
\begin{equation}
g_{ab}=e^{-2\alpha}\gJ_{ab},
\quad
g^{ab}=e^{2\alpha}\gJ^{ab},
\quad
\sqrt{-g}=e^{-4\alpha}\sqrt{-\gJ},
\label{eq:conf_basic_action}
\end{equation}
the Ricci scalar transforms as
\begin{equation}
R=e^{2\alpha}\left[\RJ+6\BoxJ\alpha-6\,\gJ^{ab}\nJ_a\alpha\,\nJ_b\alpha\right].
\label{eq:R_conf_direct}
\end{equation}
Substituting Eqs.~\eqref{eq:conf_basic_action} and \eqref{eq:R_conf_direct} into Eq.~\eqref{eq:general_action} yields
\begin{align}
S&=\frac{1}{2\kappa^2}\int d^4x\sqrt{-\gJ}\,
\Bigg[
\Afun e^{-2\alpha}\RJ
+6\Afun e^{-2\alpha}\BoxJ\alpha
\nonumber\\
&
-6\Afun e^{-2\alpha}(\nJ\alpha)^2
-\Bfun e^{-2\alpha}\gJ^{ab}\nJ_a\Phi\nJ_b\Phi
\nonumber\\
&
-2\ell^{-2}\Vfun e^{-4\alpha}
\Bigg]
+S_m[\gJ_{ab},\chi].
\label{eq:pre_ibp_direct}
\end{align}
Using Eq.~\eqref{eq:Psi_def} and Eq.~\eqref{eq:Uinv_def}, this becomes
\begin{align}
S&=\frac{1}{2\kappa^2}\int d^4x\sqrt{-\gJ}\,
\Bigg[
\Psi\,\RJ
+6\Psi\,\BoxJ\alpha
-6\Psi\,(\nJ\alpha)^2
\nonumber\\
&
-\Psi\,\frac{\Bfun}{\Afun}\,\gJ^{ab}\nJ_a\Phi\nJ_b\Phi
-2\ell^{-2}\VJ
\Bigg]
+S_m[\gJ_{ab},\chi].
\label{eq:direct_before_ibp}
\end{align}

We now integrate the $\BoxJ\alpha$ term by parts. Up to the usual total divergence,
\begin{equation}
\int d^4x\sqrt{-\gJ}\,\Psi\,\BoxJ\alpha
=
-\int d^4x\sqrt{-\gJ}\,\gJ^{ab}\nJ_a\Psi\,\nJ_b\alpha.
\label{eq:ibp_alpha}
\end{equation}
Since
\begin{equation}
\nJ_a\Psi=\Psi'\,\nJ_a\Phi,
\quad
\frac{\Psi'}{\Psi}=\frac{\Afun'}{\Afun}-2\alpha',
\quad
\nJ_a\alpha=\alpha'\,\nJ_a\Phi,
\label{eq:Psi_alpha_derivs}
\end{equation}
the action becomes
\begin{align}
S&=\frac{1}{2\kappa^2}\int d^4x\sqrt{-\gJ}\,
\Bigg[
\Psi\,\RJ-2\ell^{-2}\VJ
\nonumber\\
&-\Psi
\left(
\frac{\Bfun}{\Afun}
+6\frac{\Afun'}{\Afun}\alpha'
-6(\alpha')^2
\right)
(\nJ\Phi)^2
\Bigg]
+S_m[\gJ_{ab},\chi],
\label{eq:kinetic_pre_omega}
\end{align}
where $(\nJ\Phi)^2\equiv \gJ^{ab}\nJ_a\Phi\nJ_b\Phi$.

At this stage it is natural to rewrite the kinetic sector in terms of the invariant Brans-Dicke function. Recall that, in the invariant formalism, the quotient of derivatives of invariants is again an invariant, and using the notation and results of~\cite{Jarv:2014hma} one has
\begin{equation}
\Ifive\equiv \left(\frac{\Ione'}{2\Ione\Ithree'}\right)^2.
\label{eq:I5_def_action}
\end{equation}
Using $\Ione=e^{2\alpha}/\Afun$ and
\begin{equation}
(\Ithree')^2=\frac{2\Afun\Bfun+3(\Afun')^2}{4\Afun^2},
\label{eq:I3prime_action}
\end{equation}
one finds
\begin{equation}
\Ifive
=
\frac{\left(2\alpha'-\Afun'/\Afun\right)^2}
{2\Bfun/\Afun+3(\Afun'/\Afun)^2}.
\label{eq:I5_explicit_action}
\end{equation}
This motivates the invariant Brans-Dicke coupling function
\begin{equation}
\omegaI(\Psi)\equiv \frac{1}{2\Ifive}-\frac{3}{2},
\label{eq:omegainv_def_action}
\end{equation}
which, after a short algebraic manipulation, becomes
\begin{equation}
\omegaI(\Psi)
=
\frac{\displaystyle
\frac{\Bfun}{\Afun}
+6\frac{\Afun'}{\Afun}\alpha'
-6(\alpha')^2}
{\displaystyle
\left(\frac{\Afun'}{\Afun}-2\alpha'\right)^2}.
\label{eq:omegainv_explicit_action}
\end{equation}
On the other hand, Eq.~\eqref{eq:Psi_alpha_derivs} implies
\begin{equation}
(\nJ\Psi)^2
=
\Psi^2
\left(\frac{\Afun'}{\Afun}-2\alpha'\right)^2
(\nJ\Phi)^2.
\label{eq:dPsi2_action}
\end{equation}
Combining Eqs.~\eqref{eq:omegainv_explicit_action} and \eqref{eq:dPsi2_action}, the kinetic term in Eq.~\eqref{eq:kinetic_pre_omega} becomes
\begin{equation}
-\Psi
\left(
\frac{\Bfun}{\Afun}
+6\frac{\Afun'}{\Afun}\alpha'
-6(\alpha')^2
\right)
(\nJ\Phi)^2
=
-\frac{\omegaI(\Psi)}{\Psi}(\nJ\Psi)^2.
\label{eq:kinetic_rewrite_action}
\end{equation}
We therefore obtain
\begin{align}
S&=\frac{1}{2\kappa^2}\int d^4x\sqrt{-\gJ}\,
\Bigg[
\Psi\,\RJ
-\frac{\omegaI(\Psi)}{\Psi}\,
\gJ^{ab}\nJ_a\Psi\nJ_b\Psi
\nonumber\\
&
-2\ell^{-2}\VJ(\Psi)
\Bigg]
+S_m[\gJ_{ab},\chi].
\label{eq:Jordan_invariant_action}
\end{align}

Equation~\eqref{eq:Jordan_invariant_action} is the Jordan-like invariant form of the theory. Every quantity appearing in it is invariant under frame transformations, yet the action has the exact Brans--Dicke--Bergmann--Wagoner (BDBW) structure. Several checks are immediate. In a Jordan parametrization with \(\alpha=0\), one has \(\gJ_{ab}=g_{ab}\), \(\Psi=\Afun\), and Eq.~\eqref{eq:Jordan_invariant_action} reduces to the usual Jordan-frame BDBW action. In a canonical Einstein parametrization with \(\Afun=1\), the same invariant metric is represented as \(\gJ_{ab}=e^{2\alpha_{\rm E}}g^{\rm E}_{ab}\), so the nonminimal matter coupling is carried by the conformal factor. Finally, when \(\Psi\) is constant, the Hessian sector that appears in the field equations below vanishes, and the dissipative imperfect-fluid pieces are absent, as expected in the general-relativistic limit.
For completeness, let us note that the same result can be recovered starting instead from the Einstein-like invariant action~\eqref{eq:einvariant_action}. Indeed, using $\gJ_{ab}=\Ione\,\eh_{ab}$, together with the standard conformal transformation formulas for the volume element, inverse metric, and Ricci scalar, one arrives at Eq.~\eqref{eq:Jordan_invariant_action} after the same integration-by-parts step. The direct derivation above is nevertheless conceptually preferable here, since it makes explicit that the Jordan-like invariant representation follows directly from the original scalar-tensor action once the invariant matter metric is chosen as the basic variable.

\section{Invariant field equations in the Jordan-like representation}
\label{sec:field_eqs}

Varying the action \eqref{eq:Jordan_invariant_action} with respect to $\gJ^{ab}$ yields
\begin{align}
\tilde G_{ab}
&=\frac{\kappa^2}{\Psi}\,T^{(m)}_{ab}
+\frac{\omegaI}{\Psi^2}\left(\nJ_a\Psi\,\nJ_b\Psi-\frac{1}{2}\gJ_{ab}(\nJ\Psi)^2\right)
\nonumber\\
&\quad+\frac{1}{\Psi}\left(\nJ_a\nJ_b\Psi-\gJ_{ab}\BoxJ\Psi\right)-\frac{\ell^{-2}\VJ}{\Psi}\,\gJ_{ab},
\label{eq:metric_eq_tilde}
\end{align}
where the matter stress tensor is defined in the usual way from the minimally coupled matter action,
\begin{equation}
T^{(m)}_{ab}\equiv -\frac{2}{\sqrt{-\gJ}}\frac{\delta S_m}{\delta \gJ^{ab}},
\qquad
\nJ^b T^{(m)}_{ab}=0.
\label{eq:matter_tensor}
\end{equation}
The scalar equation is
\begin{equation}
(2\omegaI+3)\BoxJ\Psi
=-\dv{\omegaI}{\Psi}(\nJ\Psi)^2+2\ell^{-2}\bigg(\Psi\dv{\VJ}{\Psi}-2\VJ\bigg)+\kappa^2 T^{(m)},
\label{eq:scalar_eq_tilde}
\end{equation}
with $T^{(m)}\equiv \gJ^{ab}T^{(m)}_{ab}$.

Equation~\eqref{eq:metric_eq_tilde} has precisely the same structural content as ordinary Jordan-frame scalar-tensor gravity: the nonminimal-coupling/Hessian sector survives, and matter is minimally coupled to the metric used to build geometry. This immediately suggests that the effective-fluid decomposition will once again be imperfect.

To make this explicit, write Eq.~\eqref{eq:metric_eq_tilde} as
\begin{equation}
\tilde G_{ab}=\kappa^2\left(\frac{T^{(m)}_{ab}}{\Psi}+T^{(g)}_{ab}\right),
\label{eq:einstein_like_tilde}
\end{equation}
where
\begin{align}
T^{(g)}_{ab}
&\equiv\frac{1}{\kappa^2\Psi}\Bigg[\nJ_a\nJ_b\Psi-\gJ_{ab}\BoxJ\Psi-\ell^{-2}\VJ\,\gJ_{ab}
\nonumber\\
&+\frac{\omegaI}{\Psi}\left(\nJ_a\Psi\,\nJ_b\Psi-\frac{1}{2}\gJ_{ab}(\nJ\Psi)^2\right)\Bigg].
\label{eq:Tg_tilde}
\end{align}
Because of the Hessian contribution, $T^{(g)}_{ab}$ is not the stress tensor of a minimally coupled scalar field. It is therefore not expected to be a perfect fluid in a generic frame.

\section{$1+3$ decomposition and imperfect-fluid structure}
\label{sec:1plus3}

Let $u^a$ be an arbitrary unit timelike congruence with respect to $\gJ_{ab}$, so that $u^a u_a=-1$. Define the projector
\begin{equation}
h_{ab}=\gJ_{ab}+u_a u_b,
\end{equation}
and the standard kinematical split
\begin{equation}
\nJ_a u_b=-a_bu_a+\frac13\The h_{ab}+\sigma_{ab}+\omega_{ab}.
\end{equation}
We follow the standard $1+3$ covariant formalism, in which the congruence $u^a$ defines a family of local observers and its integral curves represent their flow lines~\cite{Tsagas:2007yx}. The spacetime derivative may then be resolved into a derivative along the flow and a derivative projected onto the observers' instantaneous rest spaces. Accordingly, for a scalar $f$, the overdot introduced below is the directional derivative along the integral curves of $u^a$, whereas $\DD_a f$ is the spatial derivative orthogonal to $u^a$. Thus $\dot f$ measures the evolution of $f$ seen by the observers moving with the congruence, while $\DD_a f$ measures the spatial inhomogeneity of $f$ in their local rest frame.

For any scalar $f$ we set
\begin{equation}
\dot f\equiv u^a\nJ_a f,
\qquad
\DD_a f\equiv h_a{}^b\nJ_b f,
\qquad
\DD^2 f\equiv \DD_a\DD^a f.
\end{equation}
In particular, $\DD_a f=0$ means that $f$ has no spatial gradient in the local rest frame of the congruence, although it may still evolve along the flow through $\dot f\neq0$.

We also fix our index notation explicitly. For any rank-two tensor $A_{ab}$, round and square brackets denote symmetrization and antisymmetrization, respectively,
\begin{equation}
A_{(ab)}\equiv\frac12\left(A_{ab}+A_{ba}\right),
\qquad
A_{[ab]}\equiv\frac12\left(A_{ab}-A_{ba}\right).
\end{equation}
Angle brackets denote the projected, symmetric and trace-free (PSTF) part with respect to $u^a$,
\begin{equation}
A_{\langle ab\rangle}\equiv\left(h_{(a}{}^{c}h_{b)}{}^{d}-\frac13 h_{ab}h^{cd}\right)A_{cd}.
\label{eq:PSTF_def}
\end{equation}
Thus the angle brackets used below are distinct from the antisymmetrization denoted by square brackets.
With the previous definitions in place, one has then the following identities:
\begin{align}
\nJ_a f&=-\dot f\,u_a+\DD_a f,
\\
\BoxJ f&=-\ddot f-\The\dot f+\DD^2 f+a^a\DD_a f,
\\
h_a{}^b u^c\nJ_c\nJ_b f&=\DD_a\dot f-\left(\frac13\The h_a{}^b+\sigma_a{}^b+\omega_a{}^b\right)\DD_b f
\\
h_a{}^b u^c\nJ_c\DD_b f &= \DD_a\dot f +\dot f\,a_a \nonumber \\
&-\left( \frac13\The h_a{}^b+\sigma_a{}^b+\omega_a{}^b \right)\DD_b f 
 .
\label{eq:keyidentity}
\end{align}

We now decompose the geometric tensor \eqref{eq:Tg_tilde} as
\begin{equation}
T^{(g)}_{ab}=\rho_g u_a u_b+p_g h_{ab}+2q^{(g)}_{(a}u_{b)}+\pi^{(g)}_{ab},
\end{equation}
with the usual definitions
\begin{align}
\rho_g&\equiv u^au^bT^{(g)}_{ab},
\qquad
p_g\equiv \frac13 h^{ab}T^{(g)}_{ab},
\\
q^{(g)}_a&\equiv -h_a{}^c u^d T^{(g)}_{cd},
\qquad
\pi^{(g)}_{ab}\equiv h_{\langle a}{}^c h_{b\rangle}{}^d T^{(g)}_{cd}.
\end{align}
A direct calculation gives
\begin{align}
\rho_g
&=\frac{1}{\kappa^2\Psi}\left[-\The\dot\Psi+\DD^2\Psi+\frac{\omegaI}{2\Psi}\left(\dot\Psi^2+\DD_a\Psi\,\DD^a\Psi\right)\right.\nonumber\\
&\left.+\ell^{-2}\VJ\right],
\label{eq:rho_tilde}
\\
p_g
&=\frac{1}{\kappa^2\Psi}\Bigg[\ddot\Psi+\frac23\The\dot\Psi-\frac23\DD^2\Psi-a^a\DD_a\Psi
\nonumber\\
&+\frac{\omegaI}{2\Psi}\dot\Psi^2-\frac{\omegaI}{6\Psi}\DD_a\Psi\,\DD^a\Psi-\ell^{-2}\VJ\Bigg],
\label{eq:p_tilde}
\\
q^{(g)}_a
&=\frac{1}{\kappa^2\Psi}\left[\dot\Psi\,a_a-h_a{}^b u^c\nJ_c\DD_b\Psi-\frac{\omegaI}{\Psi}\dot\Psi\,\DD_a\Psi\right]
\nonumber\\
&=\frac{1}{\kappa^2\Psi}\Bigg[-\DD_a\dot\Psi+\left(\sigma_a{}^b+\omega_a{}^b+\frac13\The h_a{}^b\right)\DD_b\Psi
\nonumber\\
&-\frac{\omegaI}{\Psi}\dot\Psi\,\DD_a\Psi\Bigg],
\label{eq:q_tilde}
\\
\pi^{(g)}_{ab}
&=\frac{1}{\kappa^2\Psi}\left[\DD_{\langle a}\DD_{b\rangle}\Psi-\dot\Psi\,\sigma_{ab}+\frac{\omegaI}{\Psi}\DD_{\langle a}\Psi\,\DD_{b\rangle}\Psi\right].
\label{eq:pi_tilde}
\end{align}
Equations~\eqref{eq:q_tilde} and \eqref{eq:pi_tilde} already establish the main point: the Jordan-like invariant representation supports a \emph{genuine imperfect fluid}. Heat flux and anisotropic stress vanish only on special branches. In particular, the Hessian term produces the same imperfect-fluid structure familiar from the ordinary Jordan-frame formulation of scalar-tensor gravity~\cite{Faraoni:2021jri,Giardino:2023ygc}.

\section{Scalar-comoving frame and invariant thermodynamic identifications}
\label{sec:thermo}

Assume now that the gradient of $\Psi$ is timelike and choose the $\Psi$-comoving congruence
\begin{equation}
u_a=\frac{\nJ_a\Psi}{\sqrt{-(\nJ\Psi)^2}}.
\label{eq:Psi_comoving}
\end{equation}
In this frame one has
\begin{equation}
\DD_a\Psi=0,
\end{equation}
which simplifies Eqs.~\eqref{eq:q_tilde} and \eqref{eq:pi_tilde} to
\begin{equation}
q^{(g)}_a\Big|_{\Psi}=\frac{\dot\Psi}{\kappa^2\Psi}\,a_a,
\label{eq:q_comoving}
\end{equation}
and
\begin{equation}
\pi^{(g)}_{ab}\Big|_{\Psi}=-\frac{\dot\Psi}{\kappa^2\Psi}\,\sigma_{ab}.
\label{eq:pi_comoving}
\end{equation}

Before doing the matching with the Eckart thermodynamics, it is useful to make the invariant character of the quantity entering the heat flux explicit. Define the norm of the invariant coupling
gradient by
\begin{equation}
\mathcal N_{\mathrm J}
\equiv
\sqrt{
-\gJ^{ab}\nJ_a\Psi\,\nJ_b\Psi
}.
\label{eq:NJ_def}
\end{equation}
Because both $\gJ_{ab}$ and $\Psi$ are invariant under the combined
Weyl rescalings and scalar-field reparametrizations,
$\mathcal N_{\mathrm J}$ is an invariant scalar. With the orientation
chosen in Eq.~\eqref{eq:Psi_comoving}, one has
\begin{equation}
\dot\Psi
=
u^a\nJ_a\Psi
=
-\mathcal N_{\mathrm J}.
\label{eq:dotPsi_NJ}
\end{equation}
Consequently,
\begin{equation}
\chiJ
\equiv
-\frac{\dot\Psi}{\kappa^2\Psi}
=
\frac{\mathcal N_{\mathrm J}}{\kappa^2\Psi}
\label{eq:chiJ_manifestly_invariant}
\end{equation}
is manifestly invariant~\cite{Jarv:2014hma}. It is therefore not merely an expression that assumes the familiar Jordan-frame form in a particular parametrization, but a scalar constructed entirely from invariant Jordan-like variables.

In terms of the invariant scalar $\chiJ$, Eqs.~\eqref{eq:q_comoving}
and \eqref{eq:pi_comoving} become
\begin{equation}
q^{(g)}_a\Big|_{\Psi}
=
-\chiJ a_a,
\label{eq:q_comoving_chi}
\end{equation}
and
\begin{equation}
\pi^{(g)}_{ab}\Big|_{\Psi}
=
\chiJ\sigma_{ab}.
\label{eq:pi_comoving_chi}
\end{equation}

Comparing Eq.~\eqref{eq:q_comoving} with Eckart's constitutive law,
\begin{equation}
q_a^{\rm Eckart}=-K\left(\DD_a T+Ta_a\right),
\label{eq:Eckart_q}
\end{equation}
we find in the scalar-comoving frame~\cite{Faraoni:2021jri}
\begin{equation}
\DD_a T_{\mathrm J}=0,
\qquad
K_{\mathrm J}T_{\mathrm J}=\chiJ\equiv -\frac{\dot\Psi}{\kappa^2\Psi}.
\label{eq:KTJ}
\end{equation}

The evolution of the thermodynamic quantity $\chiJ=K_{\mathrm J}T_{\mathrm J}$
also makes explicit how the matter sector can influence whether the
scalar--tensor gravitational sector is effectively ``heated'' or ``cooled.''
Strictly speaking, the Eckart matching determines the product
$K_{\mathrm J}T_{\mathrm J}$ rather than $T_{\mathrm J}$ separately;
therefore the following terminology refers to the evolution of $\chiJ$.
A separate evolution law for $T_{\mathrm J}$ would require additional
constitutive information specifying $K_{\mathrm J}$.

In the $\Psi$-comoving frame, $\DD_a\Psi=0$ and therefore
$\BoxJ\Psi=-\ddot\Psi-\The\dot\Psi$ and
$(\nJ\Psi)^2=-\dot\Psi^2$. Combining the scalar equation
\eqref{eq:scalar_eq_tilde} with the definition
\eqref{eq:chiJ_manifestly_invariant} gives
\begin{align}
\dot{\chi}_{\mathrm J}
&=-\The\chi_{\mathrm J}
+\frac{\dot\Psi^{\,2}}{\kappa^2\Psi^2}
\nonumber\\
&+\frac{1}{\kappa^2\Psi\left(2\omegaI+3\right)}
\left[\dv{\omegaI}{\Psi}\dot\Psi^{\,2}+2\ell^{-2}\left(\Psi\dv{\VJ}{\Psi}-2\VJ\right)\nonumber \right.\\
&\left.+\kappa^2 T^{(m)}\right].
\label{eq:chiJ_evolution}
\end{align}
Thus $\dot{\chi}_{\mathrm J}>0$ and $\dot{\chi}_{\mathrm J}<0$ correspond, respectively, to an increase and a decrease of the effective scalar--tensor thermal variable. Matter enters directly through its trace $T^{(m)}$ and indirectly through the geometry, in particular through the expansion scalar $\The$. For a perfect-fluid matter component,
\begin{equation}
T^{(m)}=-\rho_m+3p_m,
\end{equation}
so that a trace-free component such as radiation does not source the scalar directly, whereas nonconformal matter generally does. During an epoch in which ordinary matter is strongly diluted, the direct $T^{(m)}$ contribution becomes negligible, but $\chi_{\mathrm J}$ can still evolve because of the expansion, the scalar potential, and the field dependence of $\omegaI$. Therefore there is no universal matter-independent tendency toward heating or cooling: the sign of $\dot{\chi}_{\mathrm J}$ is controlled by the competition among these terms.

Likewise, comparison of Eq.~\eqref{eq:pi_comoving} with
\begin{equation}
\pi_{ab}^{\rm Eckart}=-2\eta\sigma_{ab}
\label{eq:Eckart_pi}
\end{equation}
shows that
\begin{equation}
\eta_{\mathrm J}=\frac{\dot\Psi}{2\kappa^2\Psi}=-\frac{1}{2}K_{\mathrm J}T_{\mathrm J}.
\label{eq:etaJ}
\end{equation}
Thus the standard first-order thermodynamic interpretation of scalar-tensor gravity is recovered, now expressed entirely in invariant Jordan-like variables.

A bulk-viscous identification is more delicate away from symmetry reduction, because the isotropic pressure \eqref{eq:p_tilde} is not uniquely split into ``viscous'' and ``non-viscous'' pieces in a generic spacetime. In a spatially homogeneous and isotropic sector, however, one has
\begin{equation}
a_a=\sigma_{ab}=\omega_{ab}=\DD_a\Psi=0,
\qquad
\The=3H,
\end{equation}
and the scalar equation \eqref{eq:scalar_eq_tilde} can be used to eliminate $\ddot\Psi$ from \eqref{eq:p_tilde}. The resulting FLRW pressure contains a unique term linear in $H$,
\begin{equation}
p_g^{\rm FLRW}=p_{\rm nonvisc}^{\rm FLRW}-\frac{H\dot\Psi}{\kappa^2\Psi},
\end{equation}
so that the natural cosmological identification is
\begin{equation}
\Pi_{\mathrm J}^{\rm(FLRW)}=-\zeta_{\mathrm J}\The,
\qquad
\zeta_{\mathrm J}=\frac{\dot\Psi}{3\kappa^2\Psi}
=-\frac{1}{3}K_{\mathrm J}T_{\mathrm J}.
\label{eq:zetaJ}
\end{equation}
This is the invariant Jordan-like counterpart of the FLRW bulk-viscous channel discussed in the scalar-tensor thermodynamics literature.

The physical meaning of these results is immediate. The Einstein-like invariant representation hides the nonminimal-coupling structure by absorbing it into the metric, thereby turning the scalar sector into a perfect fluid. By contrast, the Jordan-like invariant representation leaves the nonminimal-coupling/Hessian sector explicit in the field equations. The same underlying conformal class then admits an effective dissipative fluid and, with it, the corresponding temperature-based thermodynamic interpretation.

\section{Comparison with the Einstein-like invariant representation}
\label{sec:comparison}

The usefulness of the Jordan-like representation is that the thermodynamic quantities are written in invariant variables while retaining the matter-coupled geometry. This makes Eqs.~\eqref{eq:KTJ}--\eqref{eq:zetaJ} directly comparable across conformal parametrizations: the same quantities may be evaluated in Jordan variables, Einstein variables, or any other parametrization without changing their invariant meaning. The construction therefore provides a frame-invariant version of the imperfect-fluid dictionary usually formulated in Jordan-frame variables.

It is also useful to isolate what changes when one passes from the Einstein-like invariant metric \eqref{eq:ghat_def} to the Jordan-like invariant metric \eqref{eq:gJ_def}. The change is not a change of conformal class, nor a change of scalar--tensor dynamics. It is a change in the invariant representation used to organize the metric, scalar field, and matter coupling. The main differences between the two choices are summarized in Table~\ref{tab:representations}.

\begin{widetext}
\begin{center}
\begin{table}[t]
\caption{Einstein-like and Jordan-like invariant representations.}
\label{tab:representations}
\begin{ruledtabular}
\begin{tabular}{ccc}
Quantity & Einstein-like representation & Jordan-like representation \\
\hline
Invariant metric & $\eh_{ab}=\Afun g_{ab}$ & $\gJ_{ab}=e^{2\alpha}g_{ab}$ \\
Matter coupling & $S_m[\Ione\eh_{ab},\chi]$ & $S_m[\gJ_{ab},\chi]$ \\
Scalar coupling & Minimal & Nonminimal, via $\Psi\RJ$ \\
Action & Eq.~\eqref{eq:einvariant_action} & Eq.~\eqref{eq:Jordan_invariant_action} \\
Effective fluid & Perfect & Imperfect \\
Thermodynamic variable & Chemical potential & $K_{\rm J}T_{\rm J}$, Eq.~\eqref{eq:KTJ} \\
\end{tabular}
\end{ruledtabular}
\end{table}
\end{center}
\end{widetext}

\subsection{Einstein-like invariant representation}

In the Einstein-like representation, the gravitational part of the action takes the minimally coupled form \eqref{eq:einvariant_action}. The scalar sector is therefore organized as a canonical minimally coupled scalar field. Its stress tensor has the perfect-fluid form, so that the heat flux and anisotropic stress vanish in the corresponding scalar-comoving description. In this representation, the temperature-based imperfect-fluid variables are therefore absent, and the departure from general relativity is naturally described instead by the chemical-potential or diffusive variable emphasized in Ref.~\cite{Jarv:2026dbb}. This is the same structural reason why minimally coupled scalar fields in Einstein gravity do not generate the Eckart-type imperfect-fluid thermodynamics discussed in Jordan-frame scalar-tensor gravity~\cite{Faraoni:2022ixc}.

\subsection{Jordan-like invariant representation}

In the Jordan-like representation, the same theory is written in the invariant BDBW form \eqref{eq:Jordan_invariant_action}. Matter is minimally coupled to $\gJ_{ab}$, while the metric field equations take the form \eqref{eq:metric_eq_tilde}. The essential difference from the Einstein-like representation is the explicit Hessian sector in Eq.~\eqref{eq:metric_eq_tilde}. When this sector is included in the effective stress tensor \eqref{eq:Tg_tilde}, the scalar contribution is no longer organized as a minimally coupled perfect fluid.

This is made explicit by the decomposition in Sec.~\ref{sec:1plus3}. In a generic congruence, the heat flux and anisotropic stress are given by Eqs.~\eqref{eq:q_tilde} and \eqref{eq:pi_tilde}, and need not vanish. In the $\Psi$-comoving frame they reduce to Eqs.~\eqref{eq:q_comoving} and \eqref{eq:pi_comoving}, from which the identifications \eqref{eq:KTJ} and \eqref{eq:etaJ} follow by comparison with Eckart's constitutive relations. In homogeneous and isotropic sectors the same representation also gives the bulk-viscous channel \eqref{eq:zetaJ}. Thus the Jordan-like invariant representation recovers the temperature-based imperfect-fluid interpretation familiar from the Jordan-frame literature~\cite{Faraoni:2021jri,Giardino:2023ygc,Faraoni:2025alq}.

\subsection{Conceptual consequence}

The comparison shows that the two invariant representations give different effective-fluid organizations of the same underlying scalar-tensor theory. In the Einstein-like representation, the nonminimal-coupling structure is absorbed into the metric and the scalar sector is naturally described as a minimally coupled perfect fluid. In the Jordan-like representation, the matter metric is used as the invariant metric and the nonminimal-coupling structure remains explicit in the field equations, leading instead to an effective imperfect fluid.

The vanishing of the scalar-sector temperature is therefore a property of the Einstein-like organization of the invariant theory, rather than of the invariant formalism independently of representation choice. The Jordan-like metric \eqref{eq:gJ_def} is equally invariant and leads to the nonzero imperfect-fluid quantities displayed in Eqs.~\eqref{eq:q_tilde}-\eqref{eq:zetaJ}. The two descriptions are not contradictory: they are distinct invariant organizations of the same dynamics, adapted to different choices of metric representation. In this sense, the invariant formalism clarifies rather than removes the representation dependence of the effective thermodynamic interpretation.

\section{Discussion and conclusions}
\label{sec:discussion}

We have shown that the invariant formalism of scalar-tensor gravity admits a Jordan-like representation in addition to the Einstein-like one. The central object is the invariant matter metric \eqref{eq:gJ_def}, together with the invariant scalar variable \eqref{eq:Psi_def}. Written in these variables, the action takes the Brans-Dicke-Bergmann-Wagoner form \eqref{eq:Jordan_invariant_action}, with invariant functions $\VJ(\Psi)$ and $\omegaI(\Psi)$. Thus the construction remains fully invariant, but the nonminimal-coupling structure is organized differently from in the Einstein-like representation.

The consequence for the effective-fluid description follows directly from the field equations. In Eq.~\eqref{eq:metric_eq_tilde}, the Hessian contribution is explicit, and the corresponding effective stress tensor is given by Eq.~\eqref{eq:Tg_tilde}. This term is responsible for the imperfect-fluid pieces in the decomposition of $T^{(g)}_{ab}$. Indeed, Eqs.~\eqref{eq:q_tilde} and \eqref{eq:pi_tilde} show that, in a generic congruence, the effective scalar sector has nonvanishing heat flux and anisotropic stress. Therefore, in the Jordan-like invariant representation, the invariant scalar sector is not naturally organized as a perfect fluid.

In the $\Psi$-comoving frame, the imperfect-fluid structure reduces to Eqs.~\eqref{eq:q_comoving} and \eqref{eq:pi_comoving}. Comparison with Eckart's constitutive relations then gives the invariant identifications \eqref{eq:KTJ} and \eqref{eq:etaJ}. In homogeneous and isotropic sectors, the same representation also yields the FLRW bulk channel \eqref{eq:zetaJ}. With the conventions used here, the sign of this bulk coefficient follows directly from $\Theta=3H$, the pressure term preceding Eq.~\eqref{eq:zetaJ}, and the definition \eqref{eq:KTJ}.

These results clarify the scope of the vanishing-temperature conclusion obtained in the Einstein-like invariant representation. That conclusion is correct within the representation defined by Eq.~\eqref{eq:ghat_def}, where the scalar sector is minimally coupled and therefore has the perfect-fluid form familiar from minimally coupled scalar fields. However, it is not forced by the invariant formalism alone. The Jordan-like invariant representation \eqref{eq:gJ_def} is equally invariant and leads instead to the imperfect-fluid organization displayed in Eqs.~\eqref{eq:q_tilde}-\eqref{eq:pi_tilde}. The difference is therefore not in the underlying scalar-tensor dynamics, but in the invariant representation used to package the metric, scalar field, and matter coupling.

This also leaves open a useful interpretational question. The Einstein-like representation is distinguished by minimal gravitational dynamics, whereas the Jordan-like representation is distinguished by minimal matter coupling. The invariant formalism by itself does not select between these two organizations. Rather, it makes clear that different invariant representations can support different effective thermodynamic descriptions of the same conformal class: a perfect-fluid, zero-temperature description in the Einstein-like representation, and a temperature-based imperfect-fluid description in the Jordan-like one.

We have not attempted to assign observational preference to either invariant representation. Observable predictions are, of course, determined by the full scalar--tensor dynamics together with the matter coupling. The point established here is more limited: the effective thermodynamic organization of the scalar sector is not unique at the level of invariant representations. Which organization is most useful depends on the question being asked. The Einstein-like representation is adapted to minimal gravitational dynamics, while the Jordan-like representation is adapted to the matter-coupled geometry and to the imperfect-fluid decomposition.

Several extensions would be natural. One could apply the same construction to multiscalar-tensor gravity, where the invariant field-space structure may lead to a richer imperfect-fluid decomposition~\cite{Pereira:2026hpi}. It would also be useful to formulate the entropy current and entropy production directly in the Jordan-like invariant variables, allowing a more detailed comparison between the temperature-based and chemical-potential-based descriptions. Finally, the role of perturbations and matter observables should be examined in order to understand whether these different invariant organizations lead to distinct practical advantages in cosmological or astrophysical applications.
\begin{acknowledgments}
We thank the anonymous referee for the valuable suggestions provided, which have greatly contributed to enhancing the quality of this manuscript.
This work was supported by Fundação para a Ciência e a Tecnologia (FCT) through national funds under the research grant UID/04434/2025 (DOI 10.54499/UID/04434/2025).
\end{acknowledgments}

\bibliographystyle{apsrev4-2}
\bibliography{invariant_jordan_thermo_refs}

\end{document}